\documentclass[twocolumn]{aastex631}

\usepackage{psfig}
\usepackage{graphicx}	

\usepackage{subfigure}

\shorttitle{JuMBOs from photoerosion of fragmenting cores}
\shortauthors{J. L. Diamond \& R. J. Parker}

\begin{document}

\title{Formation of Jupiter-Mass Binary Objects through photoerosion of fragmenting cores}

\correspondingauthor{Richard Parker}
\email{R.Parker@sheffield.ac.uk}

\author{Jessica L. Diamond}
\affiliation{Astrophysics Research Cluster, School of Mathematical and Physical Sciences, The University of Sheffield, Hicks Building, Hounsfield Road, Sheffield, S3 7RH, UK}

\author[0000-0002-1474-7848]{Richard J. Parker}
\altaffiliation{Royal Society Dorothy Hodgkin Fellow}
\affiliation{Astrophysics Research Cluster, School of Mathematical and Physical Sciences, The University of Sheffield, Hicks Building, Hounsfield Road, Sheffield, S3 7RH, UK}

\label{firstpage}

\begin{abstract}
The recent discovery of tens of Jupiter-mass binary objects (JuMBOs) in the Orion Nebula Cluster with the James Webb Space Telescope has intensified the debate on the origin of free-floating planetary mass objects within star-forming regions. The JuMBOs have masses below the opacity limit for fragmentation, but have very wide separations (10s -- 100s\,au), suggesting that they did not form in a similar manner to other substellar mass binaries. Here, we propose that the theory of photoerosion of prestellar cores by Lyman continuum radiation from massive stars could explain the JuMBOs in the ONC. We find that for a range of gas densities the final substellar mass is comfortably within the JuMBO mass range, and that the separations of the JuMBOs are consistent with those of more massive (G- and A-type) binaries, that would have formed from the fragmentation of the cores had they not been photoeroded. The photoerosion mechanism is most effective within the H{\small{II}} region(s) driven by the massive star(s). The majority of the observed JuMBOs lie outside of these regions in the ONC, but may have formed within them and then subsequently migrated due to dynamical evolution. 
\end{abstract}

\keywords{star forming regions (1565), massive stars (732), star formation (1569), brown dwarfs (185), free-floating planets (549), H\,{\small II} regions (694)}

\section{Introduction}

Understanding the formation mechanism(s) of brown dwarfs and planetary-mass objects is one of the current outstanding problems in astrophysics. It is difficult to explain the formation of planets more than several Jupiter masses ($M_{\rm Jup}$) via the core-accretion theory of planet formation \citep{Pollack96,Chabrier14}, and it is also difficult to explain the formation of significant numbers of free-floating planetary mass objects via collapse and fragmentation, as the opacity limit for fragmentation reaches a minimum at around 10\,$M_{\rm Jup}$ \citep{Rees76,Padoan04,Bate12,Hennebelle13,Clark21}.

The discovery  of 540 free-floating Jupiter-mass objects in the Orion Nebula Cluster with the James Webb Space Telescope \citep{Pearson23} was therefore unexpected, especially given so many of these objects (42) are apparently in binary systems with both components of the binary in the mass range 0.7 -- 13\,M$_{\rm Jup}$ (hence the moniker `Jupiter-mass binary object' -- or `JuMBO' for short).

Furthermore, the fraction of these systems in binaries is much higher \citep[at around 9\,per cent,][]{Pearson23} than for other, slightly more massive substellar objects over a similar separation range (10s -- 100s\,au), which appears to be inconsistent with the previously observed trend of decreasing multiplicity fraction with decreasing primary mass \citep{Duchene13b}.

Whilst there are many different formation theories for brown dwarfs \citep[e.g.][]{Reipurth01,Whitworth04,Gahm07,Goodwin07b,Bate09,Haworth15,Mathew21}, most do not predict that the brown dwarfs would form in a binary system, and the fraction of planetary mass objects in binaries in the \citet{Pearson23} sample exceeds the binary fraction for more massive brown dwarf--brown dwarf binaries in the Galactic field \citep{Basri06,Burgasser07,Thies07}. Furthermore, the separation distribution on the JuMBOs is skewed towards significantly higher values (28 -- 384\,au) than the brown dwarf--brown dwarf binaries in the field \citep[$\sim$4\,au,][]{Burgasser07}, and in fact is more similar to the peak separations of higher-mass binary systems \citep[e.g. K-, F- and G-type, i.e.\,\,0.5 -- 1.5\,M$_\odot$ or A-type, i.e.\,\,1.5 -- 3.0\,M$_\odot$, which peak at $\sim 50$\,au and $\sim 300$\,au, respectively,][]{Raghavan10,DeRosa14}.

\begin{table*}
\caption{Stellar and substellar binary properties which we compare to the properties of the JuMBOs. We show the type of the primary mass $m_p$, the main sequence mass range this corresponds to, the binary fraction $f_{\rm bin}$, the mean separation $\bar{a}$, and the mean (${\rm log}\,\bar{a}$) and variance ($\sigma_{{\rm log}\,\bar{a}}$) of the log-normal fits to these distributions. The final column inidcates the color of each corresponding line in Fig.~\ref{jumbo_sepdist}.}
\begin{center}
\begin{tabular}{cccccccc}
\hline 
Type & Primary mass & $f_{\rm bin}$ & $\bar{a}$ & ${\rm log}\,\bar{a}$ & $\sigma_{{\rm log}\,\bar{a}}$ & Ref. & Color\\
\hline
BD &  $0.02 < m_p/$M$_\odot \leq 0.08$ & 0.15 & 4.6\,au & 0.66 & 0.4 & \citet{Burgasser07,Thies07} & orange \\
\hline
M- & $0.08 < m_p/$M$_\odot \leq 0.45$ & 0.34 & 16\,au & 1.20 & 0.80 & \citet{Bergfors10,Janson12} & blue \\
\hline 
G- & $0.8 < m_p/$M$_\odot \leq 1.2$ & 0.46 & 50\,au & 1.70 & 1.68 & \citet{Raghavan10} & red\\
\hline
A- & $1.5 < m_p/$M$_\odot \leq 3.0$ & 0.48 & 389\,au & 2.59 & 0.79 & \citet{DeRosa14} & green\\
\hline
\end{tabular}
\end{center}
\label{field_props}
\end{table*}

Whilst \citet{Pearson23} note the apparent difficulty in explaining the formation of such low-mass binary objects \citep[though several authors have proposed various dynamical mechanisms for their formation, e.g.][]{Lazzoni24,Zwart24,Wang24}, the environment of the ONC -- with its massive stars driving H{\small{II}} regions \citep{Odell17,Kang17,Odell20,Habart24} -- is conducive to forming low-mass objects via the photoerosion of pre-stellar cores \citep{Whitworth04}.

In the \citet{Whitworth04} theory \citep[see also][]{Dyson68,Kahn69,Hester96,Gahm07}, Lyman continuum radiation from massive stars drives an ionisation shock front into the prestellar core, compressing the inner layers whilst simultaneously evaporating the outer layers. The net effect is a very efficient formation of a substellar-mass object. Whilst \citet{Whitworth04} describes the formation of a single object, the multiplicity of protostars is high \citep[e.g.][]{Chen13} and so the formation of a substellar system via photoerosion might act on a primordial binary system.

In this paper we use the theory in \citet{Whitworth04} to calculate the mass of a substellar object formed via photoerosion of a pre-stellar core by radiation from massive stars. For each core that could form a substellar object(s), we then calculate the likely primary mass of the binary system that would have formed had the core not been photoeroded.

The paper is organised as follows. In Section~\ref{methods} we outline the theory and describe the properties of the various types of binary systems we will compare our results to. We present our results in Section~\ref{results}, and we discuss them in Section~\ref{discuss}. We conclude in Section~\ref{conclude}.

\section{Method}
\label{methods}

Our starting assumption is that the photoerosion mechanism described in \citet{Whitworth04} would act on a core that was already starting to fragment to form a binary or multiple system. We assume that the same amount of mass would be lost from the core during photoerosion, but that the final system would be a binary object, rather than a single object.

Stellar binaries are thought to predominantly form either through the turbulent fragmentation of a pre-stellar core \citep[e.g.][]{Donate04b,Goodwin07,Bate09,Bate12,Offner10}, or (less often) the fragmentation of a circumstellar disc \citep[e.g.][]{Rice05,Stamatellos09}. The timescales for the initial fragmentation are short \citep[high multiplicity fractions are found in young ($\leq$0.1\,Myr) protostellar cores, e.g.][]{Chen13,Luo22}, and we can compare this to the timescales for the core photoerosion model in \citet{Whitworth04}.

  \citet{Whitworth04} determine that the timescale for the first stage of the erosion ($t_1$, the time taken for the inward propagation wave to erode the outer core and compress the centre) is given by
  \begin{equation}
t_1 = \frac{1}{\left(2\pi Gn_0m\right)^{1/2}},
  \end{equation}
  where $m = m_p/X$, the mass associated with one hydrogen nucleus ($m_p$ is the proton mass, and $X$ the mass fraction of Hydrogen), and $n_0$ is the density of hydrogen nucleii in the H\small{II} region surrounding the core. We adopt two different hydrogen nucleii densities, $n_0 = 10^3$\,cm$^{-3}$ or $n_0 = 10^4$\,cm$^{-3}$, which gives either $t_1 \sim 1$\,Myr or $t_1 \sim 0.3$\,Myr, respectively. Both of these timescales are higher than the timescale for initial fragmentation ($\sim$0.1\,Myr), suggesting that the process of forming the would-be stellar binary has already set in motion before the photoerosion takes hold.

  However, we contend that photoerosion could still influence the final mass of the binary system. The initial fragmentation of the core into a binary system occurs after $\sim$0.1\,Myr, but in simulations \citet{Goodwin04b} and \citet{Bate09} show that a significant proportion of the binary component masses (up to several M$_\odot$) are accreted in the subsequent $\sim$0.1 -- 0.3\,Myr. Therefore, the core can fragment into a (low-mass) proto-binary system, but photoerosion will act to remove the material that would otherwise be accreted onto the binary system on a  similar, or faster, timescale than the accretion timescale. The net effect of this process is to reduce the final mass of the components of the binary.

\begin{figure}
\begin{center}
\rotatebox{270}{\includegraphics[scale=0.4]{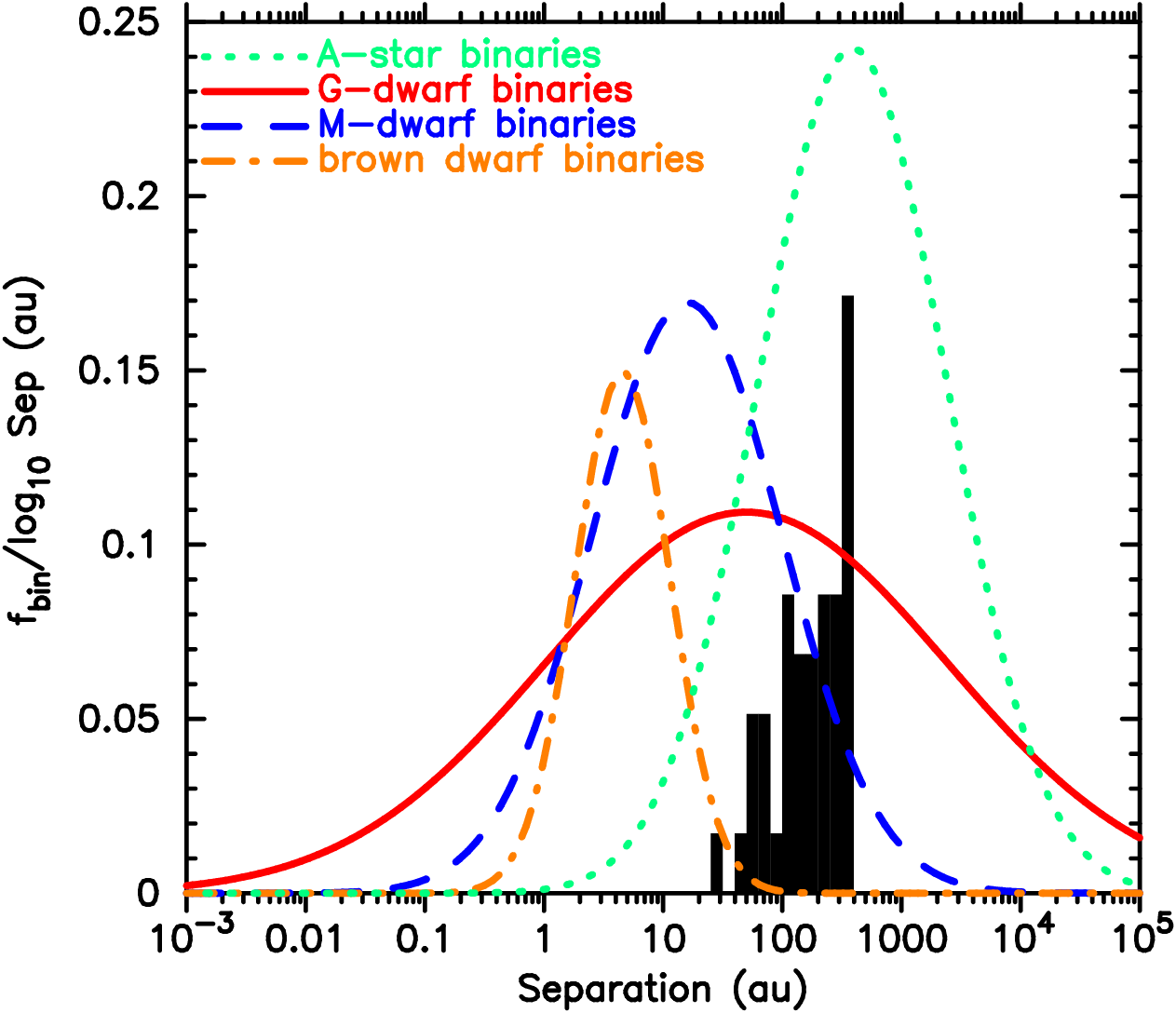}}
\caption{Separation distribution of the JuMBOs (the black histogram). For comparison, we show the Gaussian fits to the separation distributions of various stellar and substellar binary populations. The fit to the brown dwarf-brown dwarf binaries \citep{Burgasser07,Thies07} is shown by the dot-dashed orange line, the fit to the M-dwarf binaries \cite{Bergfors10,Janson12} is shown by the  dashed blue line, the fit to the G-dwarf (Solar-type) binaries \citep{Raghavan10} is shown by the solid red line, and the fit to the A-type binaries  \citep{DeRosa14} is shown by the dotted green line.}
\label{jumbo_sepdist}
  \end{center}
\end{figure}

In Fig.~\ref{jumbo_sepdist} we show the separation distribution of the JuMBOs from \citet{Pearson23} via the black histogram. For comparison, we show Gaussian fits to the separation distributions of stellar-mass binaries observed in the Galactic field (the dashed blue [M-type], solid red [G-type] and dotted green [A-type] lines), as well as the Gaussian fit to the brown dwarf--brown dwarf binary separation distribution (the dot-dashed orange line). The latter is mainly for systems in the Galactic field, although it does include several examples from nearby star-forming regions \citep{Parker11a}. A summary of these distributions (i.e.\,\,the mean separation, and variance of the Gaussian fit, and the primary mass range the fits are valid for) are given in Table~\ref{field_props}.

We also assume the core mass function is the precursor to the stellar initial mass function \citep{Alves07}, but shifted to higher masses by  a factor equal to the inverse of the star formation efficiency, $\epsilon$. We set $\epsilon = 0.333$ and draw $N_{\rm core} = 1000$ masses from a core mass function that has the same shape as the \citet{Maschberger13} stellar Initial Mass Function,
\begin{equation}
p(m) \propto \left(\frac{m}{\mu}\right)^{-\alpha}\left(1 + \left(\frac{m}{\mu}\right)^{1 - \alpha}\right)^{-\beta}.
\label{maschberger_imf}
\end{equation}
In Eqn.~\ref{maschberger_imf}  $\alpha = 2.3$ is the \citet{Salpeter55} power-law exponent for higher mass cores, and $\beta = 1.4$ describes the turnover at lower masses. As the core mass function is shifted to higher masses (by a factor of $1/\epsilon$) compared to the stellar IMF,  we adopt $\mu = 0.6$\,M$_\odot$, and sample masses in the range $M_{\rm core} = 0.3$ -- 300\,M$_\odot$.

For each core mass, $M_{\rm core}$, we use the analytical formulae from \citet{Whitworth04} to estimate the final mass of the photoeroded core, $M_3$,
\begin{equation}
M_3 \simeq \frac{2f\alpha_*GM_{\rm core}^2}{3\alpha_1a_{\rm II}^2R_{\rm HII}},
\end{equation}
where $f = 1.7$ is a factor that takes into account accretion from the remaining envelope of the core onto the low-mass object, and $G$ is the gravitational constant. $\alpha_* = 2 \times 10^{-13}$\,cm$^3$\,s$^{-1}$ is the  recombination coefficient into excited states, $\alpha_1 = 10^{-13}$\,cm$^3$\,s$^{-1}$ is the  recombination coefficient into the ground state only.  $a_{\rm II} = 10^6$cm\,s$^{-1}$ is the isothermal speed of sound in an ionised gas and $R_{\rm HII}$ is the radius of the H{\small{II}} region within which the core is being photoeroded.

The size of the H{\small{II}} region, $R_{\rm HII}$, is given by 
\begin{equation}
R_{\rm H\small{II}} = \left(\frac{3\dot{N}_{\rm Lyc}}{4\pi \alpha_* n_0^2} \right)^{1/3},
\end{equation}
where $\dot{N}_{\rm Lyc}$ is the Lyman Continuum photon rate from the massive star(s) and $n_0$ is the number density of hydrogen nuclei. Whilst \citet{Whitworth04} present results for a wide range of the parameter space, we fix  $a_{\rm II}$ but adopt two values for the density of hydrogen nuclei, $n_0 = 10^3$cm$^{-3}$ and $n_0 = 10^4$cm$^{-3}$, which for a given  $\dot{N}_{\rm Lyc}$ give two different radii for the  H{\small{II}} region(s). For lower $n_0$ the radiation penetrates further, resulting in a larger $R_{\rm HII}$, and conversely for higher $n_0$ the radiation does not penetrate as far and  $R_{\rm HII}$ is smaller.

To estimate  $\dot{N}_{\rm Lyc}$ we use the models from \citet{Schaerer97}, who provide $\dot{N}_{\rm Lyc}$ for stars of spectral types O3 -- B0.5. We take the \citet{Hillenbrand97} census of the ONC and identify six stars that have spectral types in this range. We also sum the $\dot{N}_{\rm Lyc}$ of three stars (including the most massive, $\theta^1$~Ori~C) that reside in the Trapezium system, as the  H{\small{II}} region caused by this system is likely to be larger than one driven by a single star. The different  $\dot{N}_{\rm Lyc}$ rates from the six massive stars, and the combined  $\dot{N}_{\rm Lyc}$ from the Trapezium system, are summarised in Table~\ref{star_data}.

\begin{table*}
  \caption{Properties of the most massive stars in the \citet{Hillenbrand97} census of the ONC. For cross reference with Fig.~\ref{hii_regions} in Section~\ref{results} we number the H{\small{II}} regions, then provide the massive star's name, mass, spectral type and then calculated Lyman continuum photon rate, $\dot{N}_{\rm Lyc}$, and finally the color used to represent the H{\small{II}} regions in Fig.~\ref{hii_regions}.}
  \begin{center}
    \begin{tabular}{cccccc}
      \hline
      H\small{II} region & Star & Mass & Sp.~Type & $\dot{N}_{\rm Lyc}$ & Color\\
      \hline
      1 & $\theta^1$~Ori~C & 45.7\,M$_\odot$ &  O7V & $1.22 \times 10^{49}$\,s$^{-1}$ & red \\
      2 & $\theta^2$~Ori~A & 31.2\,M$_\odot$ & O9V & $2.88 \times 10^{48}$\,s$^{-1}$ & purple \\
      3 & $\theta^1$~Ori~A & 18.9\,M$_\odot$ & O9.5V & $1.78 \times 10^{48}$\,s$^{-1}$ & green \\
      4 & $\theta^1$~Ori~D & 16.6\,M$_\odot$ & B0V & $1.05 \times 10^{48}$\,s$^{-1}$ & cyan \\
      5 & HD\,37061 & 16.3\,M$_\odot$ & B0V & $1.05 \times 10^{48}$\,s$^{-1}$ & blue \\
      6 & $\theta^2$~Ori~B & 12.0\,M$_\odot$ & B0.5V & $5.09 \times 10^{47}$\,s$^{-1}$ & orange \\
      7 & (1 + 3 + 4) & --- &  ---& $1.40 \times 10^{49}$\,s$^{-1}$ & black \\
     \hline	

\hline

\hline
    \end{tabular}
  \end{center}
  \label{star_data}
\end{table*}

We assume that the core that will be photoeroded to form a JuMBO would ordinarily form a binary or higher-order multiple system. For simplicity, and for ease of comparison with the binary statistics of stars in the Solar neighbourhood, we assume that the core will fragment (or already has fragmented) into a binary star. We assume the final \emph{system} mass of the binary if no photoerosion occurred would be $\epsilon M_c$ (the core mass multiplied by star formation efficiency within the core), and assume that the system splits into a primary mass $m_p$ and a secondary mass $m_s$ where the mass ratio $q = m_s/m_p$ is drawn from a flat distribution between zero and unity \citep{Reggiani11a,Reggiani13}. For a core of mass $M_c$, the primary mass is therefore 
\begin{equation}
m_p = \frac{\epsilon M_c}{1 + q},
\end{equation}
and the secondary mass is
\begin{equation}
  m_s = qm_p,
\end{equation}
where $q$ is randomly selected from a flat distribution  between zero and unity. 

\section{Results}
\label{results}

\begin{figure*}
  \begin{center}
\setlength{\subfigcapskip}{10pt}
\hspace*{-1.5cm}\subfigure[$n_0 = 10^3$\,cm$^{-3}$]{\label{hii_regions-a}\rotatebox{270}{\includegraphics[scale=0.35]{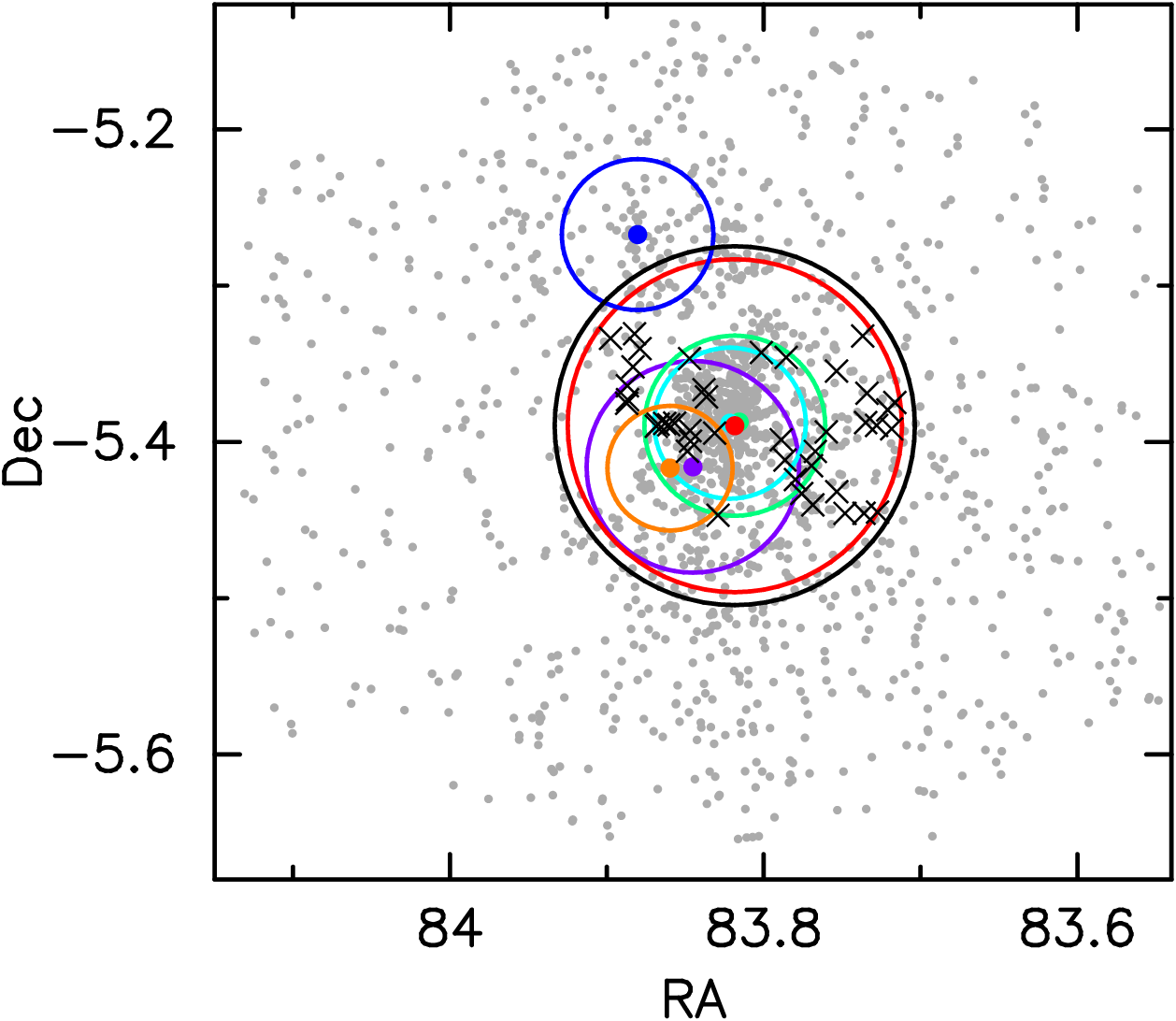}}}
\hspace*{0.3cm} 
\subfigure[$n_0 = 10^4$\,cm$^{-3}$]{\label{hii_regions-b}\rotatebox{270}{\includegraphics[scale=0.35]{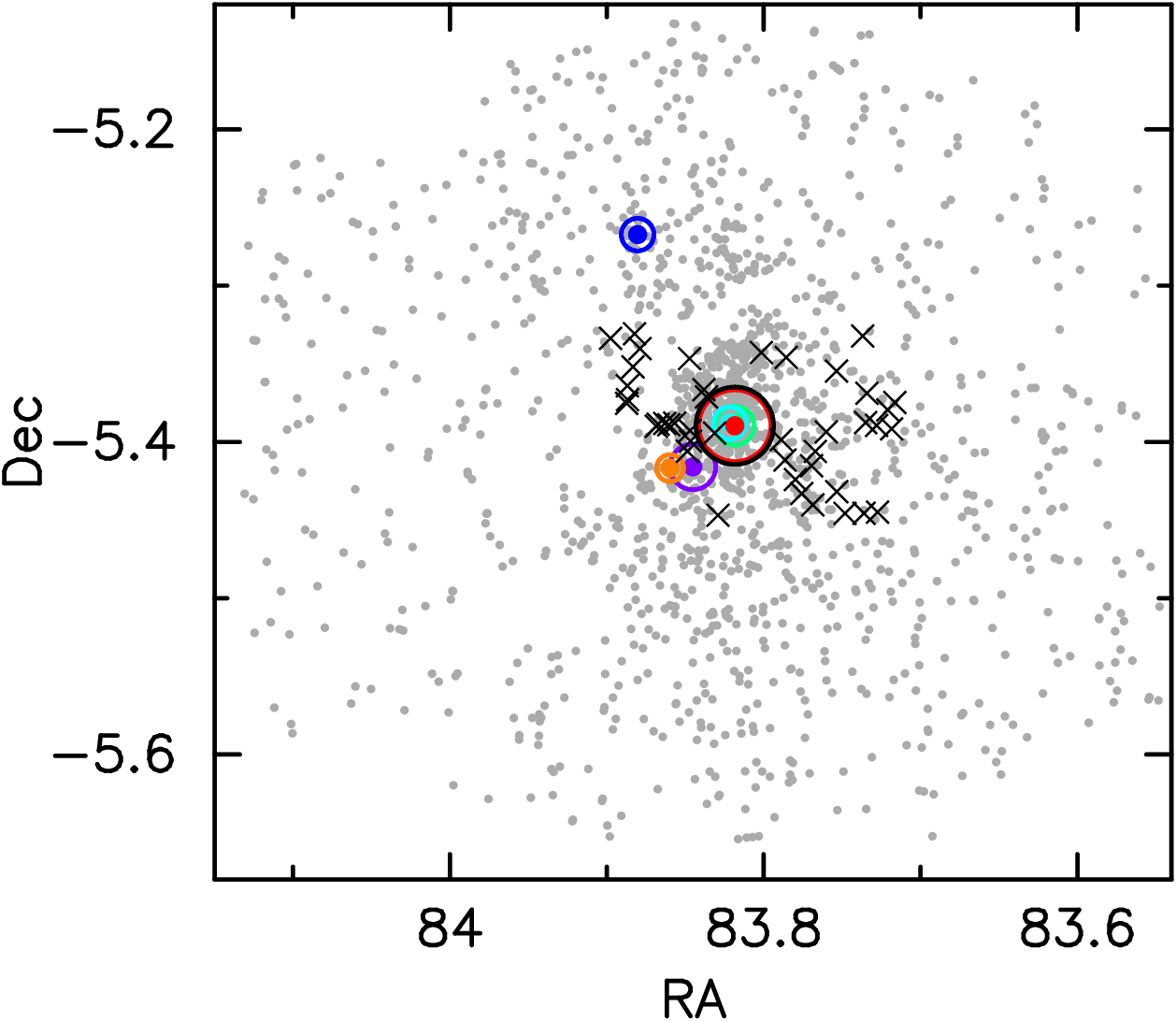}}}
\caption{Radii of H{\small{II}} regions around 6 massive stars in the ONC, as well as the H{\small{II}} region driven by the combined flux of three massive stars within the Trapezium system (the black circles). The locations of the observed JuMBOs are shown by the black crossses. The two panels show the sizes of the H{\small{II}} regions for different assumed densities of hydrogen nucleii.}
\label{hii_regions}
  \end{center}
\end{figure*}

The photoerosion mechanism described in \citet{Whitworth04} is efficient within the H{\small{II}} region around a massive star(s), and to determine the potential efficacy of this mechanism in the ONC, we plot the locations, and sizes, of the H{\small{II}} regions, as well as the positions of the JuMBOs, in Fig.~\ref{hii_regions}.

We show two versions of this plot; in Fig.~\ref{hii_regions-a} we show the position and size of the H{\small{II}} regions when we assume the density of hydrogen nuclei in the star-forming region is $n_0 = 10^3$\,cm$^{-3}$. The JuMBOs (shown by the black crosses) all lie within the H{\small{II}} region driven by  $\theta^1$~Ori~C. In contrast, when the density of hydrogen nucleii is higher ($n_0 = 10^4$\,cm$^{-3}$) the radii of the H{\small{II}} regions are much smaller (Fig.~\ref{hii_regions-b}), and the majority of JuMBOs do not lie within them.

\begin{figure*}
  \begin{center}
\setlength{\subfigcapskip}{10pt}
\hspace*{-1.5cm}\subfigure[$n_0 = 10^3$\,cm$^{-3}$]{\label{jumbo_masses-a}\rotatebox{270}{\includegraphics[scale=0.35]{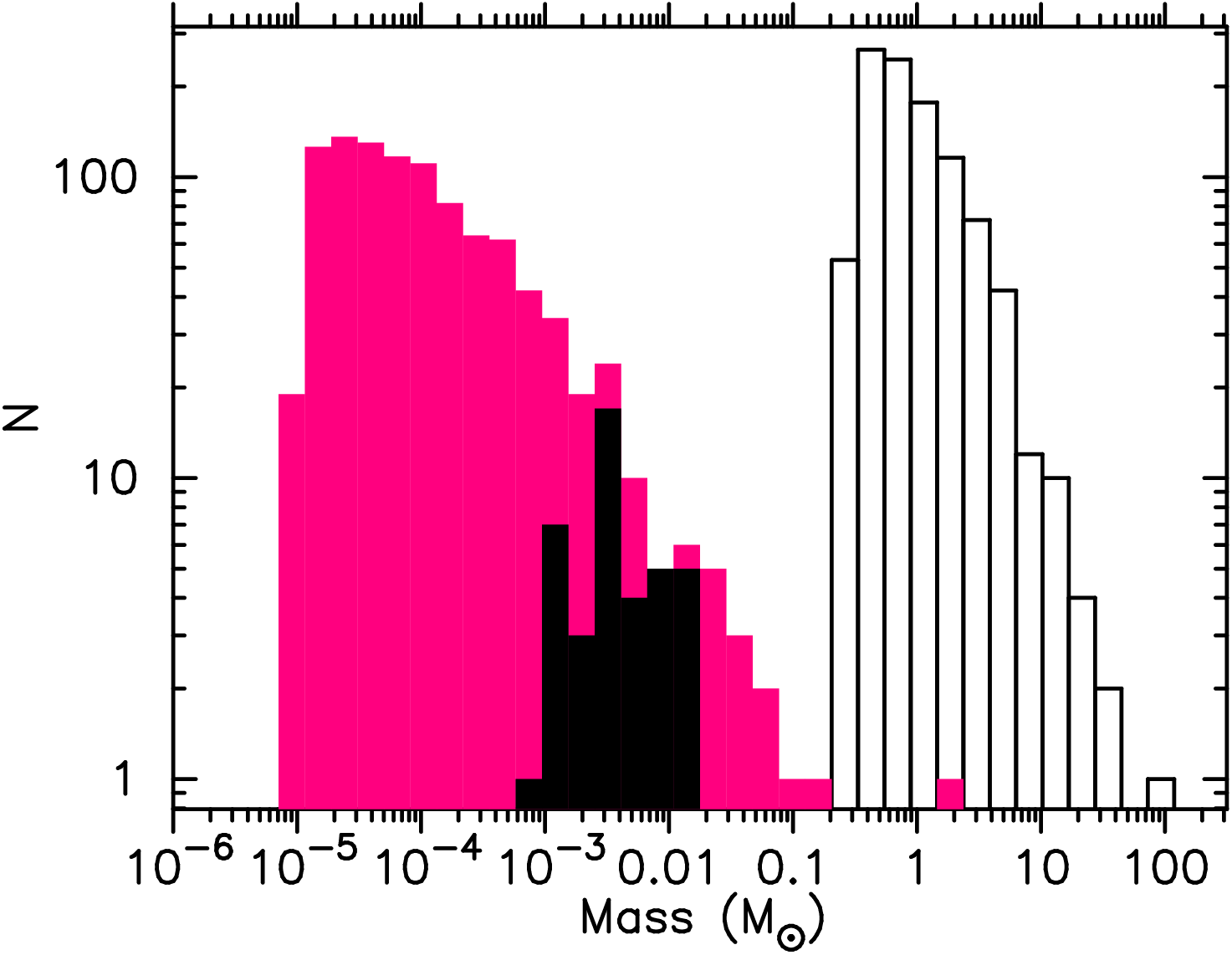}}}
\hspace*{0.3cm} 
\subfigure[$n_0 = 10^4$\,cm$^{-3}$]{\label{jumbo_masses-b}\rotatebox{270}{\includegraphics[scale=0.35]{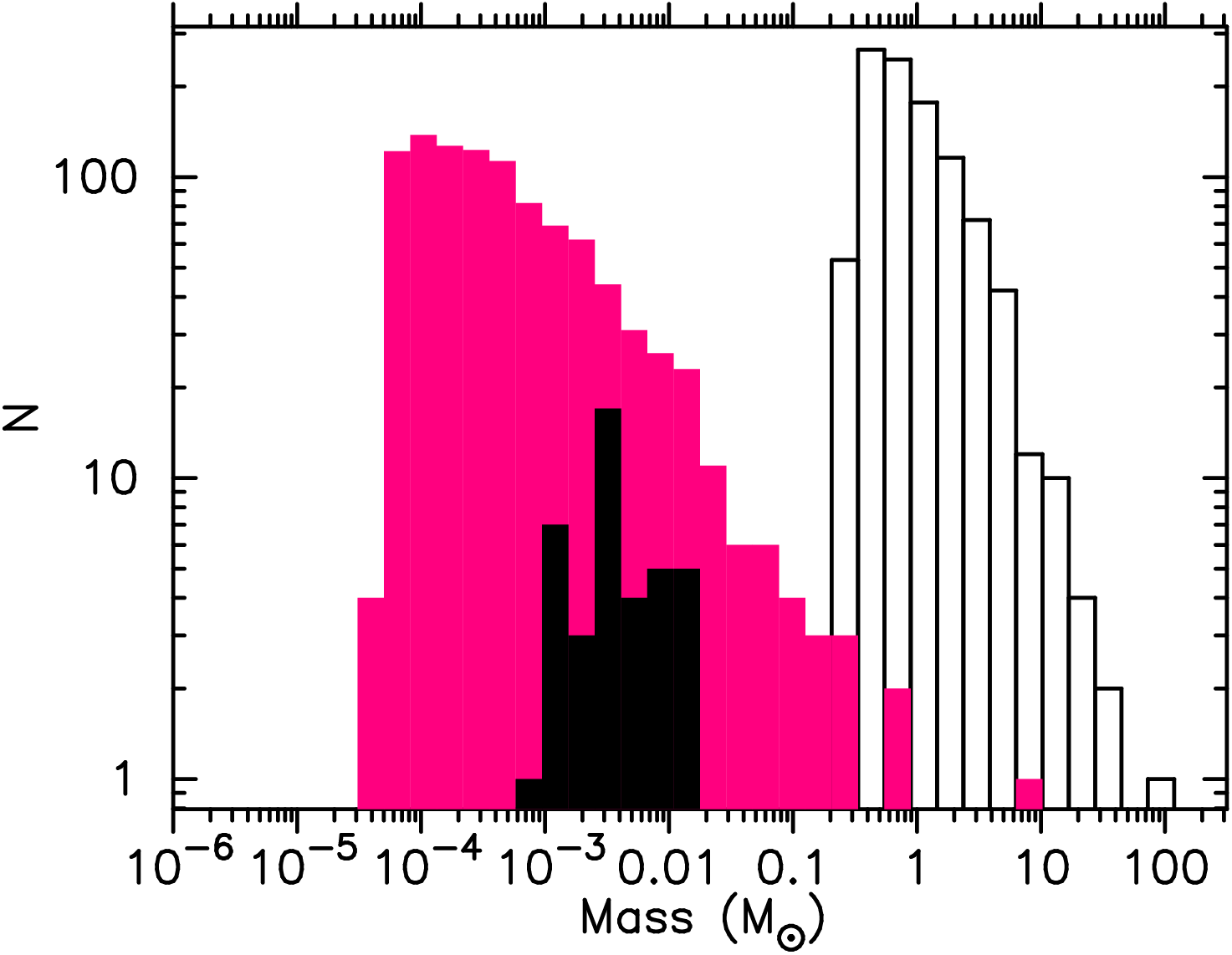}}}
\caption{Histograms of the initial core mass distributions (the open histograms), the masses, $M_3$, of the final objects following photoerosion of the cores (the pink histograms) and the observed JuMBO masses (black histograms). We show the distributions for the two different Hydrogen densities.}
\label{jumbo_masses}
  \end{center}
\end{figure*}

We now show the resultant mass distributions of the original cores, the photoeroded cores, and the observed masses of the JuMBOs, in Fig.~\ref{jumbo_masses}. Again we show the results for two different assumed Hydrogen densities (cf. H{\small{II}} region radii). The original core mass function is shown by the open histograms, and the photoeroded core masses ($M_3$) are shown by the solid pink histograms. For comparison, the JuMBO mass distribution is shown by the solid black histograms.

Irrespective of the assumed Hydrogen density, the final $M_3$ masses lie comfortably within the range of the observed JuMBO masses. In the case of the lower Hydrogen densities (larger H{\small{II}} regions, Fig.~\ref{jumbo_masses-a}) the observed JuMBOs lie towards the high-mass end of the $M_3$ masses, whereas in the case of the higher Hydrogen densities (smaller H{\small{II}} regions, Fig.~\ref{jumbo_masses-b}), the observed JuMBO masses lie within the centre of the $M_3$ mass distribution.  

\begin{figure*}
  \begin{center}
\setlength{\subfigcapskip}{10pt}
\hspace*{-1.5cm}\subfigure[$n_0 = 10^3$\,cm$^{-3}$]{\label{jumbo_ini_fin-a}\rotatebox{270}{\includegraphics[scale=0.35]{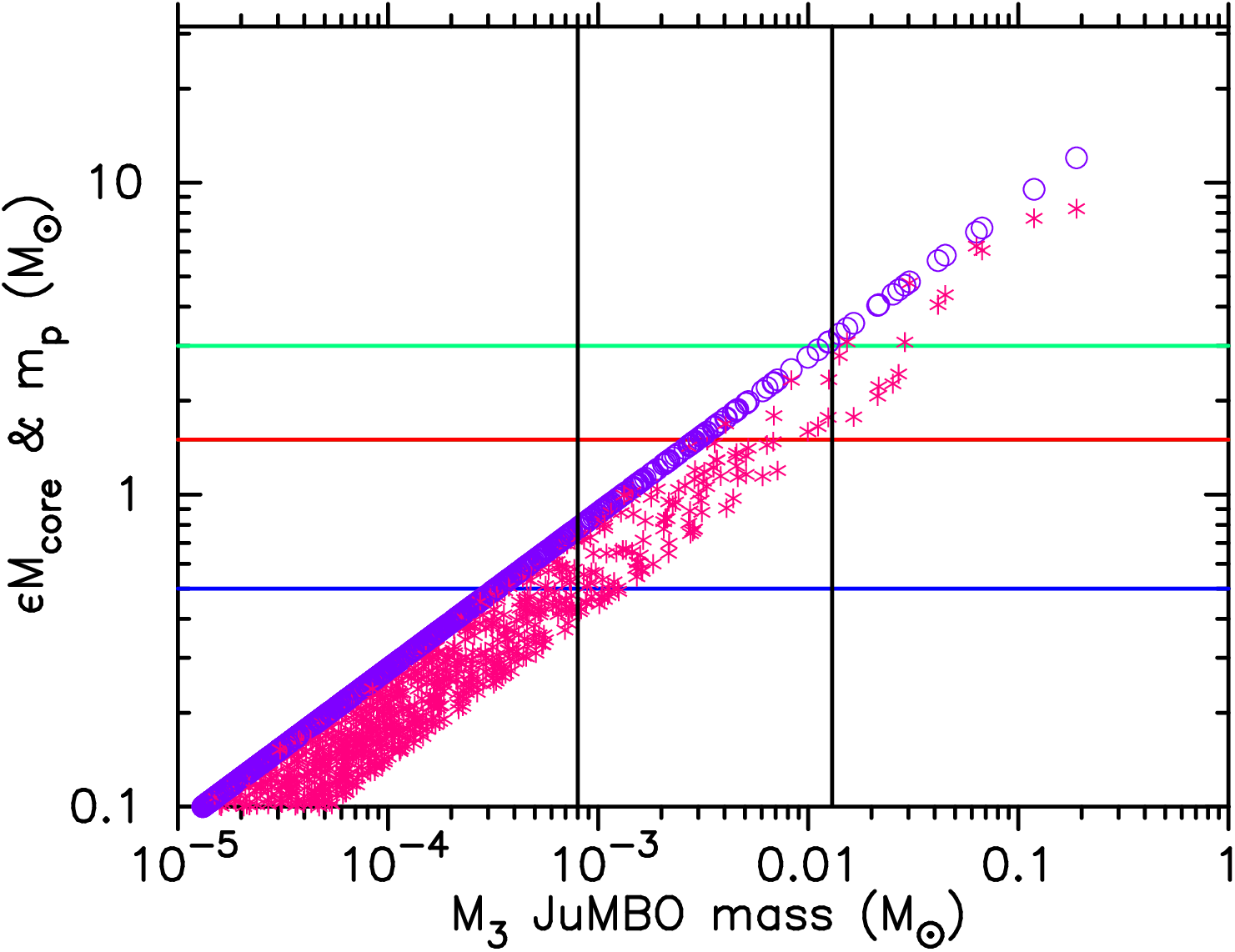}}}
\hspace*{0.3cm} 
\subfigure[$n_0 = 10^4$\,cm$^{-3}$]{\label{jumbo_ini_fin-b}\rotatebox{270}{\includegraphics[scale=0.35]{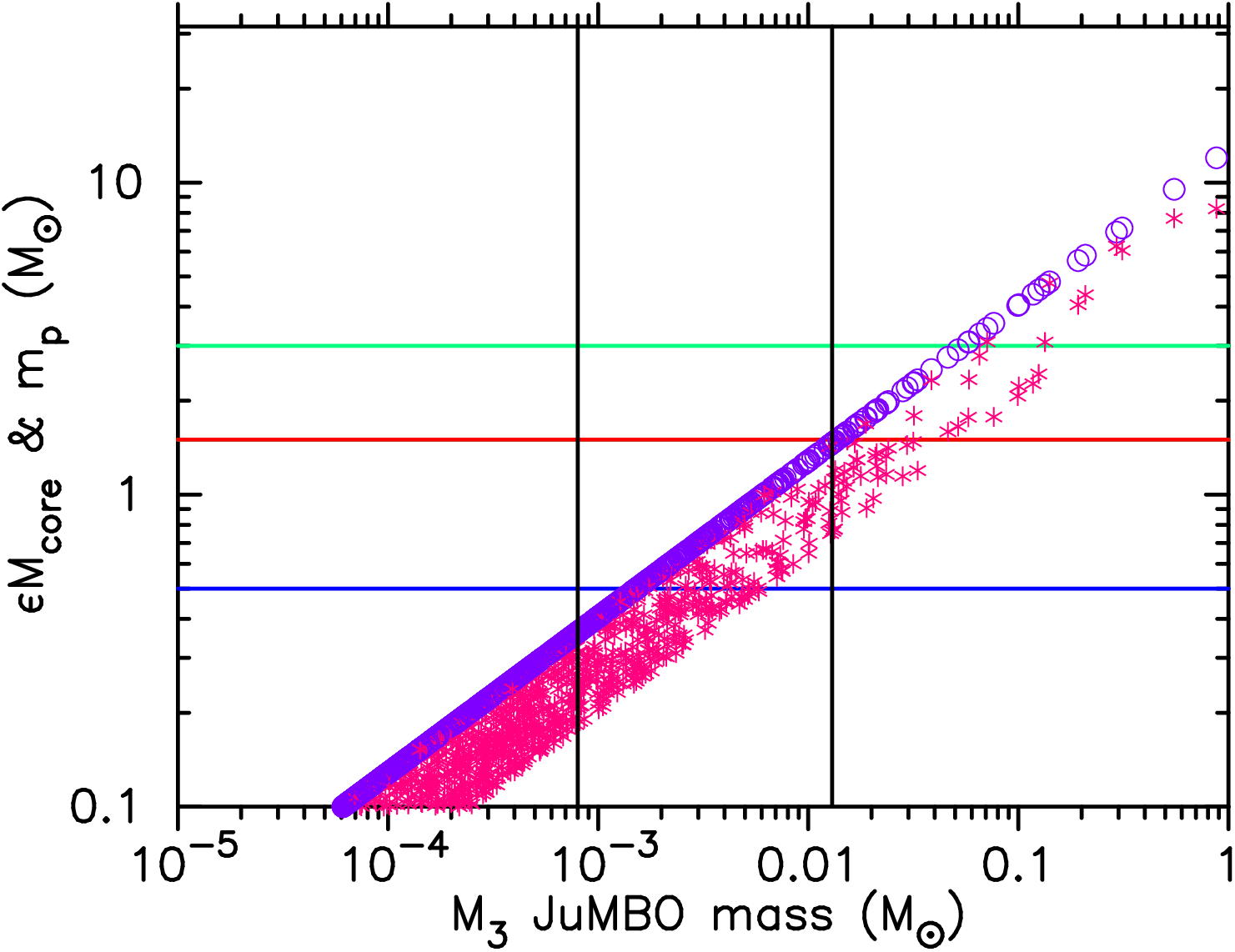}}}
\caption{Product of star formation efficiency and core mass, $\epsilon M_c$, against final JuMBO mass, $M_3$, shown by the purple circles. The primary mass of the binary, $m_p$, if it had been allowed to form without mass-loss due to photoerosion, against final JuMBO mass $M_3$, is shown by the pink asterisks. The lower and upper limits on the JuMBO masses from \citet{Pearson23} are shown by the vertical black lines, and the upper limits to the primary masses, $m_p$, of M-type (blue), G-type (red) and A-type (green) binaries are shown by the horizontal lines.}
\label{jumbo_ini_fin}
  \end{center}
\end{figure*}

Our underlying assumption is that the photoerosion mechanism would act upon a core that has already fragmented, or is in the process of fragmenting, into a multiple system. For simplicity we assume that the multiple system is a binary (although in reality higher-order systems are just as, if not more, likely). In Fig.~\ref{jumbo_ini_fin} we plot the product of the star formation efficiency and the core mass, $\epsilon M_c$ against the final JuMBO mass, $M_3$ as the purple open circles. We also plot the primary mass $m_p$, if the core  formed a binary star without being subject to photoerosion, against the final JuMBO mass, $M_3$, by the pink asterisks.

For comparison, we show the range of observed JuMBO masses by the vertical black lines, and then the upper mass limit for primary masses of M-type (blue), G-type (red) and A-type (green) binary star systems by the horizontal lines. For the lower Hydrogen densities (panel a) the majority of cores that form JuMBOs would have gone on to form G-type binary stars, whereas for the higher Hydrogen densities (panel b) the initial core masses need to be lower, and so the cores that form JuMBOs would have gone on to form a mixture of M- and G-type binary systems.

\section{Discussion}
\label{discuss}

We have demonstrated that the photoerosion formation mechanism for substellar objects can explain the masses of the JuMBOs, and their relatively high separations (28--384\,au) compared to the brown dwarf--brown dwarf binaries in the Galaxy ($<$10\,au), if they formed from a fragmenting core that would otherwise have gone on to form a more massive (Solar-type) binary or multiple system.

We draw the reader's attention to the following caveats. First, the photoerosion theory was developed to explain single object formation. Whilst there is nothing to preclude its application to binary systems, we note that the compression of the prestellar core may not be as efficient if that core is already fragmenting.

Although we have shown that the initial timescales for photoerosion of the core are longer than the fragmentation timescales, and so we might expect photoerosion to act on a core that is already forming a binary or multiple system,  the temperature of the core may be higher in the H{\small II} regions. There have been contradictory results in the literature on whether higher core temperatures result in reduced fragmentation (and therefore reduced/altered binary formation).  \citet{Offner09} note that both the Jeans length in a turbulent core, and the Toomre fragmentation criterion in a disc, are proportional to the gas temperature and we would therefore expect a higher temperature to suppress both small-scale fragmentation \citep[see also][]{Sigalotti23} and the formation of brown dwarfs \citep[see also][]{Bate12}. However, \citet{Bate12} find that the binary properties (multiplicity fraction, separation distribution) are very similar in simulations with radiative transfer compared to those without.

  \citet{Luo22} find tentative observational evidence that in the Orion star-forming complex the cores with the highest dust temperatures (attributed to external heating) have the lowest multiplicity fractions. Conversely, in a theoretical study, \citet{Guszejnov23} find that slightly more multiple systems form in simulations with higher external radiation fields. None of these authors report a significant difference in the separation distributions in binaries that form in warmer cores.

Second, the theory assumes that photoerosion only occurs within the H{\small{II}} region driven by the massive star(s). This is highly dependent on the assumed density of Hydrogen nucleii in the star-forming region, $n_0$, with higher densities leading to smaller H{\small{II}} regions, and vice versa. If we assume $n_0 = 10^3$cm$^{-3}$, then $R_{\rm HII}$ are typically $\sim 0.6$pc, and all of the observed JuMBOs lie within a hypothetical H{\small{II}} region driven by the massive stars in the Trapezium system. However, if $n_0 = 10^4$cm$^{-3}$, then $R_{\rm HII}$ are typically $\sim 0.2$pc, and the vast majority of JuMBOs would not currently reside within an H{\small{II}} region.

Observationally, \citet{Odell17} find that $\theta^1$Ori~C dominates the ionisation of the central regions of the ONC, but $\theta^2$Ori~A does dominate a more distant region, suggesting the influence of  $\theta^1$Ori~C is limited and its H{\small{II}} region is small. \citet{Habart24} use JWST observations and establish the size of the photodissociation region around $\theta^1$Ori~C as $\sim0.2$\,pc \citep[see also][]{Odell20}, and derive a Hydrogen density $n_0 \geq 10^4$cm$^{-3}$. Outside of the photodissociation region(s), observations by \citet{Weilbacher15} suggest densities in the range $3 \times 10^3 - 2 \times 10^4$\,cm$^{-3}$. Both densities are still within the range over which the \citet{Whitworth04} mechanism can operate.

If the H{\small{II}} region in the ONC is this small, this does not necessarily mean that the JuMBOs did not form within the vicinity, as they could have subsequently moved due to dynamical interactions. The ONC is one of the most dense star-forming regions in the nearby ($<$1\,kpc) Galaxy and is thought to be many dynamical timescales old \citep[e.g.][]{Furesz08,Tobin09,Allison11,DaRio17,Schoettler20,Farias20}. \citet{Allison10} show that the dynamical timescale of interest -- the \emph{local} crossing time -- in the ONC is likely to be $\leq$0.1\,Myr, meaning that for a typical velocity dispersion in a dynamically evolved system \citep[$\sim$1\,km\,s$^{-1}$, roughly 1\,pc\,Myr$^{-1}$,][]{Parker16b}, stars (and JuMBOs) can travel many pc in the 1 -- 4\,Myr old ONC \citep{Jeffries11,Reggiani11b,Beccari17}.

Furthermore, simulations that follow the photoevaporation of protoplanetary discs find a high rate of disc destruction in the ONC \citep{ConchaRamirez19,Winter19b,Parker21a},  and this is corroborated by observations \citep[e.g.][]{Henney99,Ballering23}. It is entirely feasible that cores could be photoeroded within a small H{\small{II}} region and then migrate elsewhere in the star-forming region. In these simulations, stars whose discs are destroyed by photoevaporation move in and out of the vicinity of the massive stars(s) \citep{Parker21b,Marchington22}.

 The high probability of disc photoevaporation raises a separate issue, in that photoevaporation of the outer regions of circumstellar discs could limit the material accreted from the disc onto the central star (or binary), the quantity of which is thought to be significant \citep[e.g.][]{Hartmann98,Stamatellos11}. However, even in a strong external UV radiation field such as that in the ONC, \citet{Parker21a} show that significant mass-loss from the disc occurs on timescales of 1\,Myr, which are either similar to, or longer than the initial core photoerosion timescales of $\sim0.3 - 1$\,Myr derived in \citet{Whitworth04}.  The binary is therefore likely to have assembled the majority of its mass before accretion from the disc is halted by photoevaporation of the disc, and mass-loss due to photoerosion of the outer core likely dominates over any mass lost due to photoevaporation of the disc.

If the Hydrogen density is low (and the corresponding H{\small{II}} region is large), we would expect that that there should be many more lower-mass JuMBOs awaiting discovery. Conversely, if the Hydrogen density is higher (and the H{\small{II}} region is smaller), then there should be more (as yet undiscovered) massive JuMBOs. Our analysis demonstrates that these systems would be closer in mass to M-dwarf binary systems if they were able to form without hindrance from photoerosion, and as such would have a closer separation distribution to the observed BD-BD binary systems in the field \citep[and therefore harder to distinguish from other substellar binaries,][]{Ward-Duong15}.    

\section{Conclusions}
\label{conclude}

We have applied the \citet{Whitworth04} theory of brown dwarf formation via the photoerosion of prestellar cores by Lyman continuum radiation from massive stars to explain the Jupiter-Mass Binary Objects (JuMBOs) in the ONC. The JuMBOS have component masses in the range 0.7 -- 13M$_{\rm Jup}$ and on-sky separations of 28 -- 384\,au. We assume the Lyman continuum radiation creates an H{\small{II}} region around the massive stars in the ONC, and determine both the size of the H{\small{II}} region, and the masses of the JuMBOs that could form from photoerosion of the cores. Our conclusions are the following:

(i) For reasonable values of the density of hydrogen nuclei ($10^3 - 10^4$\,cm$^{-3}$), the final masses of the photoeroded cores lie comfortably within the mass range of the observed JuMBOs.

(ii) For the lower density $10^3$\,cm$^{-3}$, the Lyman continuum radiation from the massive stars in the ONC would penetrate furthest and produce an H{\small{II}} region with a radius of $\sim$0.6\,pc. The observed JuMBOs all reside within this distance of the most massive stars in the ONC.

(iii) However, the measured density of hydrogen in the ONC is $\geq 10^4$\,cm$^{-3}$ \citep{Habart24}, which implies a photodissociation radius no larger than 0.2\,pc from the most massive stars in the ONC. The majority of the observed JuMBOs lie outside this smaller region, and at these higher densities therefore would have had to have formed within this smaller region, and then migrated out due to dynamical interactions.\\

Irrespective of the gas density and size of the H{\small{II}} region, the separations of the JuMBOs are much higher than those of substellar mass binaries (which typically peak at $\sim$4\,au, and presumably formed via a different mechanism). Instead, the JuMBOs are consistent with being the outcome of a photoeroded core that was already in the process of fragmenting to form a binary or multiple system.


\section*{Acknowledgements}

We thank the two anonymous referees for their comments and suggestions on the original manuscript. RJP acknowledges support from the Royal Society in the form of a Dorothy Hodgkin Fellowship. 



\bibliographystyle{aasjournal}  
\bibliography{general_ref}


\label{lastpage}

\end{document}